\documentclass[sigconf]{acmart}
\AtBeginDocument{%
  }

\copyrightyear{2026}
\acmYear{2026}
\setcopyright{cc}
\setcctype{by}
\acmConference[RecSys '26]{20th ACM Conference on Recommender Systems}{September 27-October 02, 2026}{Minneapolis, MN, USA}
\acmBooktitle{20th ACM Conference on Recommender Systems (RecSys '26), September 27-October 02, 2026, Minneapolis, MN, USA}
\acmDOI{10.1145/3773078.3831838}
\acmISBN{979-8-4007-2284-4/2026/09}

\usepackage{multirow}
\usepackage{tabularray}
\usepackage{pgfplots}
\usepackage{booktabs}
\usepgfplotslibrary{groupplots}
\pgfplotsset{compat=1.18}
\definecolor{curveblue}{RGB}{33,113,181}
\definecolor{curveorange}{RGB}{217,95,2}
\begin{document} %todo\sloppy

%%
%% Thtitle" commTODO has an optional parameter,
\title{PriCoRec: A Privacy-Aware Cloud–Device Collaborative Framework for Ad Recommendation under Feature Constraints}

\author{Dairui Liu}
\authornote{Both authors contributed equally to this research.}
\author{Zhongyi Lu}
\authornotemark[1]

\affiliation{%
  \institution{University College Dublin}
  \city{Dublin}
  \country{Ireland}}
\email{dairui.liu@ucd.ie}
\email{zhongyi.lu@ucd.ie}

% \author{Zhongyi Lu}
% \email{zhongyi.lu@ucd.ie}
% \affiliation{%
%   \institution{University College Dublin}
%   \country{Ireland}}

\author{Jitao Lu}
\affiliation{%
  \institution{University College Dublin}
  \city{Dublin}
  \country{Ireland}}
\email{jitao.lu@ucd.ie}

\author{Aghiles Salah}
\affiliation{%
  \institution{Huawei Ireland Research Center}
  \city{Dublin}
  \country{Ireland}}
\email{aghiles.salah@h-partners.com}

\author{Mete Sertkan}
\affiliation{%
  \institution{Huawei Ireland Research Center}
  \city{Dublin}
  \country{Ireland}}
\email{mete.sertkan@h-partners.com}

\author{Roger Zhe Li}
\affiliation{%
  \institution{Huawei Ireland Research Center}
  \city{Dublin}
  \country{Ireland}}
\email{roger.zhe.li@huawei.com}

\author{Changhong Jin}
\affiliation{%
  \institution{University College Dublin}
  \city{Dublin}
  \country{Ireland}}
\email{changhong.jin@ucd.ie}

\author{Barry Smyth}
\affiliation{%
  \institution{University College Dublin}
  \city{Dublin}
  \country{Ireland}}
\email{barry.smyth@ucd.ie}

\author{Xingsheng Guo}
\affiliation{%
  \institution{Huawei Ireland Research Center}
  \city{Dublin}
  \country{Ireland}}
\email{xingsheng.guo1@huawei-partners.com}

\author{Ruihai Dong}
\affiliation{%
  \institution{University College Dublin}
  \city{Dublin}
  \country{Ireland}}
\email{ruihai.dong@ucd.ie}

\renewcommand{\shortauthors}{Dairui Liu et al.}

\begin{abstract}
Privacy regulations increasingly restrict cloud processing of sensitive user data (e.g., age, gender), hindering traditional cloud-only recommendation models. To mitigate this challenge, we propose a \textbf{Pri}vacy-aware \textbf{Co}llaborative cloud-device ads \textbf{Rec}ommendation framework (PriCoRec) which personalizes recommendations while keeping sensitive features on-device. While separating recommendation into cloud-based and on-device stages enables privacy-aware deployment, naive splitting suffers from degraded shortlist quality and inefficient on-device inference due to limited private features. We therefore design a collaborative framework that comprises a cloud-based pre-ranking stage using cloud-accessible features, and an on-device ranking stage that locally incorporates highly personalized features. We introduce a diversity regularizer to pre-ranking to improve candidate quality.
Moreover, to control device power consumption and computational cost, we incorporate a cloud-guided training mechanism that enhances device model performance while keeping the model lightweight. Experiments demonstrate that the proposed framework maintains strong recommendation performance while keeping sensitive features on-device.
\end{abstract}



% \received{13 February 2026}

\maketitle

\section{Introduction}\label{sec:introduction}
Computational advertising relies on large-scale CTR prediction models for real-time ranking and monetization. Deep neural networks capture high-order feature interactions~\cite{DBLP:conf/ijcai/ZhangQGTH21} and can be viewed as personalized recommendation based on user preferences~\cite{DBLP:journals/ker/BaiGDXJYL25, DBLP:journals/corr/abs-2407-01712}, often involving sensitive features (e.g., age)~\cite{DBLP:conf/sigir/Feng0PMZ0Z25}. However, privacy regulations and consent constraints frequently restrict cloud access to such features, challenging model effectiveness.

While privacy-preserving techniques such as Federated Learning~\cite{DBLP:journals/tnn/SunXLHKWJC25} and Differential Privacy~\cite{DBLP:conf/kdd/McSherryM09} provide strong data protection, their deployment is often limited by cloud accessibility constraints on sensitive features, along with significant computational overhead and infrastructure requirements,  motivating a framework to better balance privacy guarantees with deployment efficiency.

% While existing privacy-preserving techniques like Federated Learning~\cite{DBLP:journals/tnn/SunXLHKWJC25} and Differential Privacy~\cite{DBLP:conf/kdd/McSherryM09} offer principled data protection, their practical deployment largely hindered by strict cloud accessibility constraints on sensitive features. These methods may introduce additional or even prohibitive system and computational overhead or require radical infrastructure overhauls, motivating a framework redesign to better balance privacy guarantees with deployment efficiency.

\begin{figure*}[t]
    \centering
    \includegraphics[width=0.7\linewidth]{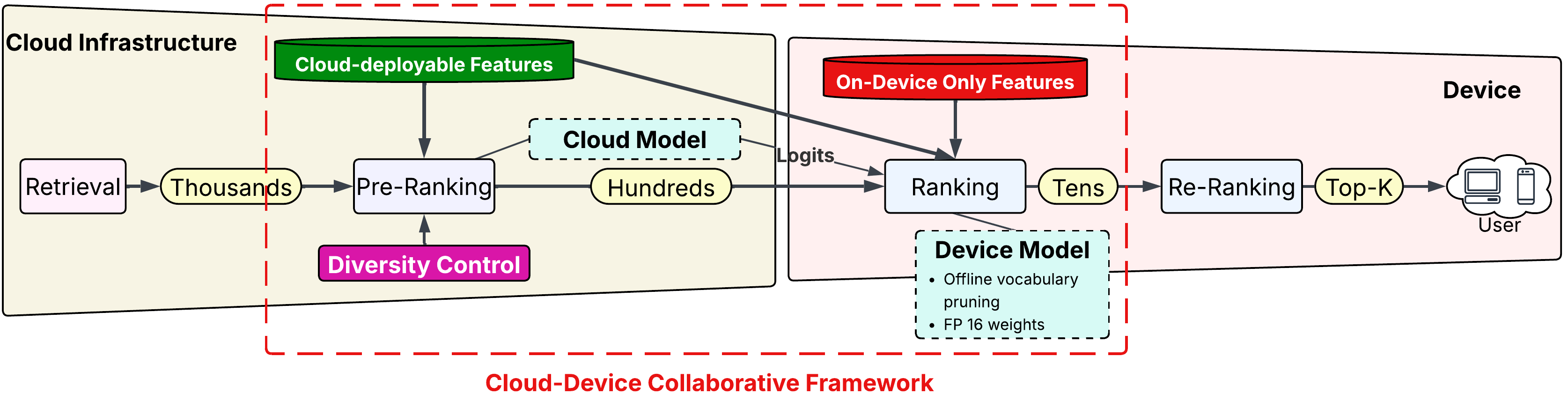}
    % \caption{Full pipeline: The red dashed rectangle highlights the proposed collaborative cloud-device framework.}
    \caption{PriCoRec's cloud--device collaboration within the four-stage recommendation pipeline.}
    \Description{A four-stage recommendation pipeline comprising retrieval, cloud pre-ranking, on-device ranking, and re-ranking. A red dashed rectangle encloses the proposed cloud pre-ranking and on-device ranking collaboration.}
    \label{fig:model_architecture}
\end{figure*}

We study large-scale CTR prediction in a cloud-based cascaded recommendation pipeline comprising \emph{retrieval}, \emph{pre-ranking}, \emph{ranking}, and \emph{re-ranking} stages~\cite{DBLP:conf/cikm/YangYBGYZ25}. While retrieval performs embedding-based filtering~\cite{DBLP:conf/recsys/FukumotoSSTSMX25}, later stages utilize neural architectures to model complex feature interactions for personalization. In standard deployments, all computation is performed in the cloud, with only the final results returned to the device~\cite{DBLP:conf/kdd/0051HQLJPLLS25}. This raises a key question: \emph{How can we redesign the ad recommendation framework to achieve personalization while ensuring user privacy?}

Moving the entire framework to the device is infeasible due to limited compute and sparse, non-iid data. Thus, we propose a collaborative cloud–device framework, moving ranking and re-ranking models to the user device to access sensitive features while keeping pre-ranking in the cloud.\footnote{Our open source code is here: \url{https://github.com/Ruixinhua/PriCoRec-RecSys2026}}In contrast to the performance-driven on-device approach~\cite{DBLP:conf/kdd/XiLWTZZZY23}, our architecture is uniquely designed to ensure privacy by isolating sensitive features from the cloud.

To address this, we propose a collaborative cloud–device framework that keeps pre-ranking in the cloud while moving ranking and re-ranking to the device, enabling local use of sensitive features. It tackles feature asymmetry through two mechanisms: (i) a DPP-inspired diversity control mechanism to improve shortlist diversity quality under restricted features, and (ii) a cloud-guided auxiliary learning strategy that enhances the lightweight on-device model without increasing inference cost.
Unlike prior cloud–device methods~\cite{DBLP:conf/kdd/XiLWTZZZY23}, which primarily focus on performance or efficiency, our framework explicitly targets privacy-preserving feature separation and its associated challenges.

Extensive experiments were conducted on three public datasets: \textbf{OpenMCC}~\cite{DBLP:conf/kdd/YaoWJHZY21}, \textbf{TaobaoAd}~\cite{DBLP:conf/ijcai/FengLSWSZY19}, and \textbf{Ali-CCP}~\cite{DBLP:conf/sigir/MaZHWHZG18}. 
The results show that our framework achieves strong prediction performance and low latency while keeping sensitive user features on-device.

% The main contributions of our work are as follows:
% \begin{itemize}
%     \item We study ad recommendation under a feature-asymmetric setting and propose a cloud–device collaborative framework that decouples cloud-side pre-ranking from device-side ranking, allowing sensitive features to remain on-device.
    
%     \item To address feature constraints, we introduce a DPP-inspired diversity regularizer to enhance candidate diversity and maintain an effective shortlist coverage. Additionally, we propose a cloud-guided auxiliary learning mechanism to improve the device-side inference performance while keeping the device model lightweight. 
%     % that leverages cloud-side predictions to improve the device model without increasing inference cost at deployment.

%     \item We conduct extensive experiments on three public datasets, demonstrating that it achieves strong CTR prediction and low latency while strictly adhering to privacy constraints.
% \end{itemize}

The main contributions of our work are as follows:
\begin{itemize}
    \item We study privacy-preserving ad recommendation under feature asymmetry and propose a cloud–device collaborative framework that separates cloud-side pre-ranking from device-side ranking, keeping sensitive features on-device.
    
    \item We introduce a DPP-inspired diversity regularizer to improve shortlist diversity under restricted features and a cloud-guided auxiliary learning strategy to enhance the lightweight device model without increasing inference cost.

    \item Extensive experiments on three public datasets demonstrate that our framework achieves strong CTR prediction and low latency while satisfying privacy constraints.
\end{itemize}

\section{Methodology}\label{sec:methodology}
As shwon in Figure~\ref{fig:model_architecture}, the focused stages are highlighted.

\subsection{Definition of Feature Groups}\label{sec:methodology:feature_groups}

We distinguish two feature groups: cloud-accessible and device-only. \textbf{Cloud-accessible} features include item features and contextual signals (e.g., request context and coarse-grained environment information) that are available at request time in the cloud without requiring access to user-sensitive attributes. \textbf{Device-only} features are sensitive user-related attributes (e.g., gender and age) and fine-grained behavioral or temporal signals, which are stored and processed locally on the user device and are not exposed to the cloud. The storage and processing of the latter feature group are restricted to be on the user device, while the former feature groups, can be deployed on either the cloud or user device.

\subsection{Cloud-Based Pre-Ranking}\label{sec:methodology:pre-ranking}
\subsubsection{Pre-Ranking Models}\label{sec:methodology:pre-ranking:models}
Using cloud-accessible features, the pre-ranking stage reduces thousands of candidates to a few hundred under strict latency constraints. While lightweight models such as LR~\cite{DBLP:conf/mm/ChenSLLH16} and DNN-based approaches~\cite{DBLP:conf/recsys/CovingtonAS16, DBLP:conf/recsys/Cheng0HSCAACCIA16} are commonly used, we adopt a more classic neural pre-ranking backbone: PNN~\cite{DBLP:conf/icdm/QuCRZYWW16}, to improve candidate recall for downstream ranking.

\subsubsection{Diversity Regularizer}\label{sec:methodology:pre-ranking:diversity_control}
To address limited cloud feature accessibility, we introduce a diversity regularizer inspired by the DPP-based algorithm of \citet{DBLP:conf/nips/ChenZZ18} as an additional objective alongside the primary ranking loss. It models item diversity through a kernel matrix capturing pairwise item similarities. By penalizing redundancy, this regularizer encourages a more diverse candidate set for subsequent on-device ranking.

The similarity between two items $i$ and $j$ is calculated as Eq.~\eqref{eqn:similarity}:
\begin{equation}
S_{i,j} = \frac{1 + \langle \tilde{I}_i, \tilde{I}_j \rangle}{2},
\label{eqn:similarity}
\end{equation}
where $<\tilde{I}_{i},\tilde{I}_{j}>$ denotes the inner product of the L2-normalized embeddings of items $i$ and $j$ derived from the pre-ranking model's embedding layer. By calculating pairwise similarities for the $n$ candidate ads of user $u$, we obtain the similarity matrix as follows
\begin{equation}
S_{R_{u}} = \begin{bmatrix}
S_{11} &\cdots  & S_{1n} \\
\vdots  &  \ddots & \vdots  \\
S_{n1} & \cdots & S_{nn}
\end{bmatrix}.
\label{eqn:matrix}
\end{equation}
Based on the similarity matrix, we define the diversity loss, $L_{div}$, as:
\begin{equation}
L_{div} = - \log det(S_{R_{u}}),
\label{eqn:div_loss}    
\end{equation}
where \emph{$\log det$} represents the log-determinant. Intuitively, a higher $\log det$ of the similarity matrix indicates greater diversity in the item list, as it reflects a larger volume spanned by the rows (or columns) of the matrix, meaning the items are less similar to each other. In this work, the recommendation score is the pCTR score. 

The total loss is the weighted sum of the base loss $L_{base}$ of the pre-ranking model and $L_{div}$. The form of $L_{base}$ depends on the chosen pre-ranking model. We adopt the pairwise ranking loss from Bayesian Personalized Ranking (BPR)~\cite{DBLP:conf/uai/RendleFGS09}, as pre-ranking focuses on ordering candidate items. For each user, one positive (clicked) item and multiple negative (non-clicked) items (1:4 random downsampling) are used to form training pairs, encouraging higher scores for positives. We apply this protocol across all pre-ranking backbones. The hyperparameter $\lambda$ is tuned within $[10^{-5}, 10^{-2}]$.
\begin{equation}
    L_{total} = L_{base} + \lambda \cdot L_{div}
\label{eqn:total_loss}
\end{equation}
By approximating the kernel log-determinant, the loss penalizes redundancy, improving diversity while preserving relevance.
\subsection{On-Device Ranking}\label{sec:methodology:re-ranking}
% \subsubsection{Cloud-Guided Training}\label{sec:methodology:re-ranking:distillation}
% We use a cloud-guided ranking framework to control the device-side model size. The pre-ranking model uses cloud-accessible features to produce relevance logits for shortlisted candidates. These logits are sent to the device together with the candidate item features. A lightweight on-device model with fewer parameters then uses the cloud logits as an auxiliary feature for ranking. To further compress the model, we apply offline vocabulary pruning to reduce the embedding dimension from 32 to 4. We also adopt FP16 weights, reducing the device model size by approximately 50\%~\cite{Rodriguez2021}.
% \subsubsection{Ranking}\label{sec:methodology:re-ranking:ranking}
% With user-sensitive features available, the on-device ranking model leverages both cloud-accessible and device-only features to refine the candidate set produced by the pre-ranking stage. The input consists of user-sensitive features, item features, and the cloud model’s relevance logit as an auxiliary feature. We adopt PNN as the backbone, which outputs a ranking score for each candidate. The higher the ranking score of a candidate item is, the better chance it will be recommended to the user. By incorporating more personalized signals during both training and inference, the model enables finer-grained discrimination, typically reducing the candidate pool by an additional order of magnitude.
\subsubsection{Cloud-Guided Ranking}\label{sec:methodology:re-ranking:distillation}

We use cloud guidance to keep the device model lightweight. The cloud pre-ranker sends each shortlisted item's relevance logit and features to the device, where a lightweight model uses the logit as an auxiliary feature instead of reproducing the larger cloud model. The cloud and device embedding dimensions are 32 and 4, respectively.

Separately, offline vocabulary remapping reduces the number of embedding-table rows, and FP16 storage reduces floating-point checkpoint size~\cite{Rodriguez2021}. These compression choices are independent and can be configured for the target device budget.
\subsubsection{Ranking}\label{sec:methodology:re-ranking:ranking}

% The on-device PNN combines cloud-accessible inputs, locally available device-only features, and the cloud relevance logit. It outputs a ranking score for every shortlisted candidate and orders the ranking-stage list. This lets the device use richer personalized signals during inference without returning any device-only feature to the cloud. The subsequent re-ranking/Top-K stage in Figure~\ref{fig:model_architecture} is outside our scope.

The on-device PNN combines cloud-accessible features, device-only features, and the cloud relevance logit to rank shortlisted candidates, enabling personalized inference without exposing device-only features to the cloud. The subsequent re-ranking/Top-K stage in Figure~\ref{fig:model_architecture} is beyond the scope of this work.

\section{Experiments}\label{sec:experiments}
% In this section, we present the experimental setup and results to validate the effectiveness of the proposed privacy-preserving cloud-device collaborative pipeline. In addition to assessing the average accuracy of users' recommendation lists from different ranking architectures, we investigate the performance of different pre-ranking models, the impact of diversity control, and how these factors influence the on-device ranking.
We evaluate the proposed cloud--device collaborative pipeline. In addition to comparing cloud pre-ranking and on-device ranking quality under fixed experimental feature boundaries, we report measured inference efficiency for OpenMCC and TaobaoAd.

\subsection{Setup}
\subsubsection{Datasets}
We conduct experiments on three large-scale industrial datasets: \textbf{OpenMCC}~\cite{DBLP:conf/kdd/YaoWJHZY21}, \textbf{TaobaoAd}~\cite{DBLP:conf/ijcai/FengLSWSZY19}, and \textbf{Ali-CCP}~\cite{DBLP:conf/sigir/MaZHWHZG18}. 
We adopt the official train/valid/test splits when available, and for datasets without predefined validation sets, we randomly split the test set into 50\% for validation and 50\% for testing. Training uses 1:4 random negative downsampling~\cite{DBLP:conf/www/HeLZNHC17}.
% To facilitate the training process based on a pairwise setup, we inherit the best practice reported in~\cite{DBLP:conf/www/HeLZNHC17} to fix the ratio between positive and negative samples at 1:4 by random negative downsampling.

\begin{table}[t]
    \centering
    \caption{Dataset Statistics.}
    \label{tab:datasets}
    \small
    \begin{tabular}{r l}
        \toprule
        \textbf{Dataset} & \textbf{Statistics} \\
        \midrule
        \multirow{4}{*}{\textbf{OpenMCC}}
         & \#Users: 5,865,593, \#Items: 381,345 \\
         & \#Interactions: 9,672,681, \#Clicks: 563,753 \\
         & \#Cloud-Accessible: 22, e.g., candidate item IDs \\
         & \#Device-Only: 5, e.g., user gender \\
        \midrule
        \multirow{4}{*}{\textbf{TaobaoAd}}
         & \#Users: 1,083,798, \#Items: 806,874 \\
         & \#Interactions: 23,341,680, \#Clicks: 1,291,950 \\
         & \#Cloud-Accessible: 11, e.g., ads group IDs \\
         & \#Device-Only: 8, e.g., user occupation \\
        \midrule
        \multirow{4}{*}{\textbf{Ali-CCP}}
         & \#Users: 356,908 \#Items: 3,168,707 \\
         & \#Interactions: 42,738,779, \#Clicks: 2,083,130 \\
         & \#Cloud-Accessible: 14, e.g., shop ID \\
         & \#Device-Only: 9, e.g., user age. \\
        \bottomrule
    \end{tabular}
\end{table}

\begin{table}[b]
    \centering
    \caption{Cloud pre-ranking with cloud-accessible features (\%); best in \textbf{bold}.}
    \Description{Cloud pre-ranking gAUC and Recall at 100 for Popularity, PNN, DP-SGD, DualRec, FedCAR, FedCIA, and PriCoRec on OpenMCC, TaobaoAd, and Ali-CCP. PriCoRec has the highest reported value for every dataset and metric.}
    \label{tab:pre_ranking}
    \small
    \begin{tabular}{lcccccc}
    \toprule
    \multirow{2}{*}{Method}
    & \multicolumn{2}{c}{OpenMCC}
    & \multicolumn{2}{c}{TaobaoAd}
    & \multicolumn{2}{c}{Ali-CCP} \\
    \cmidrule(lr){2-3}
    \cmidrule(lr){4-5}
    \cmidrule(lr){6-7}
    & gAUC & R@100 & gAUC & R@100 & gAUC & R@100 \\
    \midrule
    Popularity
    & -- & 35.82
    & -- & 23.12
    & -- & 25.95 \\

    PNN
    & 79.95 & 49.44
    & 88.69 & 67.30
    & 72.40 & 46.86 \\

    DP-SGD
    & 80.06 & 49.72
    & 89.25 & 67.85
    & 72.51 & 46.91 \\

    DualRec
    & 79.99 & 49.10
    & 88.47 & 66.74
    & 72.10 & 46.78 \\

    FedCAR
    & 79.59 & 49.61
    & 88.66 & 67.65
    & 72.82 & 46.56 \\

    FedCIA
    & 79.70 & 49.82
    & 88.59 & 67.47
    & 72.93 & 46.92 \\

    \midrule
    \textbf{PriCoRec}
    & \textbf{80.29} & \textbf{50.41}
    & \textbf{89.36} & \textbf{68.29}
    & \textbf{73.69} & \textbf{47.72} \\

    \bottomrule
    \end{tabular}
\end{table}

As described in Section~\ref{sec:methodology:feature_groups}, we categorize features as cloud-accessible or device-only and map each dataset to this unified taxonomy. Specifically, OpenMCC provides item and user features; item features and sequential behavior features are treated as cloud-accessible, while the remaining profile features are device-only. TaobaoAd organizes features into item, shopping-behavior, and user-profile groups; we treat the first two groups as cloud-accessible and the last as device-only. Ali-CCP contains item, shop, and user-behavior features; item and request-side contextual features are treated as cloud-accessible, while user-history signals are treated as device-only. The resulting statistics are summarized in Table~\ref{tab:datasets}.

\subsubsection{Baselines and Models}

We use PNN as the backbone in both stages and compare with DP-SGD~\cite{DBLP:conf/ccs/AbadiCGMMT016}, DualRec~\cite{DBLP:journals/tmc/ZhangDYLR25}, FedCAR~\cite{DBLP:journals/corr/abs-2412-11463}, and FedCIA~\cite{DBLP:conf/sigir/Han0XLGZGL25}, adapting all methods to the same feature constraints. For PriCoRec, we add the diversity objective to the cloud pre-ranker and feed its relevance logit to the device ranker. The cloud PNN uses 32-dimensional embeddings and three layers, whereas the device PNN uses 4-dimensional embeddings and two layers. Hyperparameters are selected on validation data.

\begin{table*}[t]
\centering
\caption{On-device ranking over the top-100 cloud candidates (\%); best in \textbf{bold}.}
\Description{On-device gAUC, Recall at 1, Recall at 10, nDCG at 10, and reciprocal rank for PNN, DP-SGD, DualRec, FedCAR, FedCIA, and PriCoRec on three datasets. PriCoRec has the highest reported value for every dataset and metric.}
\label{tab:on_device_ranking}
\small

\begin{tabular}{lccccccccccccccc}
\toprule
\multirow{2}{*}{\textbf{Method}}
& \multicolumn{5}{c}{\textbf{OpenMCC}}
& \multicolumn{5}{c}{\textbf{TaobaoAd}}
& \multicolumn{5}{c}{\textbf{Ali-CCP}} \\
\cmidrule(lr){2-6}
\cmidrule(lr){7-11}
\cmidrule(lr){12-16}
& \textbf{gAUC} & \textbf{R@1} & \textbf{R@10} & \textbf{N@10} & \textbf{RR}
& \textbf{gAUC} & \textbf{R@1} & \textbf{R@10} & \textbf{N@10} & \textbf{RR}
& \textbf{gAUC} & \textbf{R@1} & \textbf{R@10} & \textbf{N@10} & \textbf{RR} \\
\midrule
PNN
& 78.44 & 1.97 & 10.72 & 5.58 & 5.33
& 83.31 & 3.03 & 15.24 & 8.19 & 7.47
& 71.39 & 4.82 & 20.26 & 11.60 & 10.01 \\

DP-SGD
& 77.99 & 3.31 & 11.37 & 7.11 & 6.79
& 85.41 & 3.49 & 17.38 & 9.37 & 8.43
& 71.46 & 4.91 & 20.97 & 11.61 & 10.12 \\

DualRec
& 77.86 & 3.32 & 10.66 & 6.51 & 6.33
& 85.32 & 3.44 & 16.53 & 8.72 & 7.63
& 71.26 & 4.61 & 20.22 & 11.35 & 9.72 \\

FedCAR
& 77.71 & 3.21 & 10.17 & 6.19 & 6.13
& 85.73 & 3.39 & 16.99 & 8.81 & 7.32
& 71.08 & 4.15 & 20.46 & 11.16 & 10.05 \\

FedCIA
& 77.57 & 3.34 & 10.85 & 6.94 & 6.02
& 85.50 & 3.84 & 16.16 & 8.38 & 7.50
& 71.68 & 5.15 & 20.76 & 12.26 & 11.05 \\
\midrule
\textbf{PriCoRec}
& \textbf{79.76} & \textbf{3.96} & \textbf{13.87} & \textbf{8.12} & \textbf{7.68}
& \textbf{86.93} & \textbf{4.14} & \textbf{21.74} & \textbf{11.60} & \textbf{10.11}
& \textbf{73.30} & \textbf{7.06} & \textbf{24.10} & \textbf{16.02} & \textbf{13.43} \\
\bottomrule
\end{tabular}

\end{table*}

\subsubsection{Evaluation Protocol}
% Our experimental pipeline simulates a real-world industrial system across three cascaded stages. To ensure realism, we include a retrieval stage even though it is not the main focus of this work. We employ a DSSM model~\cite{DBLP:conf/cikm/HuangHGDAH13} to retrieve 1,000 candidate items for each request from the massive item pool. 
Although retrieval is not the focus, we evaluate the full cascade: a DSSM model~\cite{DBLP:conf/cikm/HuangHGDAH13} retrieves 1{,}000 candidates, the cloud pre-ranker retains 100, and the device model re-ranks that shortlist. We report user-grouped AUC (gAUC), Recall@K (R@K), nDCG@K (N@K), and reciprocal rank (RR). R@100 measures whether a clicked item reaches the device candidate set.
% In the current setup, Recall@1000 is 44.94\% for OpenMCC, 30.06\% for TaobaoAd, and 64.78\% for Ali-CCP.

\subsection{Pre-ranking Results}
% The cloud-side pre-ranking stage uses only cloud-accessible features to reduce 1{,}000 retrieved candidates to 100 shortlisted items for device-side ranking. At this stage, Recall@100 is especially important because it measures whether the clicked item survives into the candidate set processed on-device.
% As shown in Table~\ref{tab:pre_ranking}, PriCoRec consistently achieves the best gAUC and Recall@100 (R@100) on all three datasets. On OpenMCC, PriCoRec improves gAUC from 80.06 to 80.29 and R@100 from 49.82 to 50.41. On TaobaoAd, it improves gAUC from 89.25 to 89.36 and R@100 from 67.85 to 68.29. On Ali-CCP, it improves gAUC from 72.93 to 73.69 and R@100 from 46.92 to 47.72. These results indicate that the proposed pre-ranking strategy improves candidate quality under restricted feature access. PriCoRec also outperforms DP-SGD, DualRec, and the federated baselines (FedCAR and FedCIA) across all datasets, suggesting that explicitly compensating for missing personalized signals is more effective in this setting than directly applying off-the-shelf privacy-aware training protocols to the cloud pre-ranking stage.

The cloud pre-ranking stage uses only cloud-accessible features to reduce 1{,}000 retrieved candidates to 100 items for device-side ranking, making R@100 a key metric to measure whether a clicked item survives into the candidate set processed on-device. Table~\ref{tab:pre_ranking} shows the highest reported gAUC and R@100 for PriCoRec on all three datasets. Relative to the strongest listed comparator, gAUC rises from 80.06 to 80.29 on OpenMCC, 89.25 to 89.36 on TaobaoAd, and 72.93 to 73.69 on Ali-CCP; R@100 rises from 49.82 to 50.41, 67.85 to 68.29, and 46.92 to 47.72, respectively. The results indicate that explicitly compensating for missing personalized signals can improve shortlist quality.
In our cloud-accessible-features setting, $\lambda$ behaves as a relevance-diversity trade-off parameter rather than a monotonic knob: small non-zero values ($10^{-3}$--$10^{-2}$) give the best Recall@100, while stronger regularization can further improve Diversity@100 under the RBF kernel. We therefore tune $\lambda$ on the validation set instead of fixing it to zero or always preferring the largest value. The optimal $\lambda$ values are set to $10^{-3}$ for OpenMCC, $10^{-2}$ for TaobaoAd, and $10^{-2}$ for Ali-CCP.

buzhi\subsection{On-device Ranking Results}

The on-device ranking stage re-ranks the top-100 candidates using all feature groups, including sensitive user features, with a lightweight PNN backbone. Table~\ref{tab:on_device_ranking} shows that PriCoRec achieves the best performance across all datasets and metrics. In particular, it reaches gAUC values of 79.76 on OpenMCC, 86.93 on TaobaoAd, and 73.30 on Ali-CCP, while also achieving the best R@1, R@10, N@10, and RR values. On OpenMCC, R@10 increases from 11.37 to 13.87. On TaobaoAd, R@10 increases from 17.38 to 21.74. On Ali-CCP, R@10 increases from 20.97 to 24.10. These gains over PNN, DP-SGD, and federated baselines demonstrate that improved cloud-side shortlists enhance final on-device ranking through effective cloud-device collaboration.

\subsection{Efficiency Analysis}
We briefly discuss deployment efficiency from the perspectives of latency and model size.
\paragraph{Latency.}
% Using the current deployed configuration, the cloud-device pipeline takes 10.32\,ms in total on TaobaoAd (9.81\,ms cloud pre-ranking + 0.51\,ms on-device ranking), compared with 11.62\,ms for a pure-cloud pipeline using all feature groups. On OpenMCC, the collaborative pipeline takes 5.08\,ms in total (4.31\,ms + 0.77\,ms), compared with 4.86\,ms for the corresponding pure-cloud pipeline. Although the split pipeline introduces a small overhead on OpenMCC, both settings remain comfortably within typical industrial real-time budgets (10--50\,ms)~\cite{DBLP:conf/sigmod/KersbergenSS22}, while ensuring that sensitive features never leave the device.
The cloud--device pipeline takes 10.32\,ms on TaobaoAd (9.81\,ms cloud pre-ranking + 0.51\,ms device ranking), versus 11.62\,ms for the pure-cloud comparator. On OpenMCC it takes 5.08\,ms (4.31\,ms + 0.77\,ms), versus 4.86\,ms. Thus, splitting the cascade can improve or slightly increase latency depending on the workload, while the measured values remain within commonly reported industrial budgets~\cite{DBLP:conf/sigmod/KersbergenSS22}.

\paragraph{Model size versus ranking quality.}
Device configurations exhibit a clear size-performance trade-off: R@1 increases from 4.51 at 7.7\,MB to 5.51 at 15\,MB, 6.51 at 22.2\,MB, and 7.51 at 29.4\,MB. This trend is consistent with the expected gain from additional embedding capacity. In practice, the smaller 7.7--15\,MB models provide attractive operating points for memory- and power-constrained devices, while larger models provide higher top-rank accuracy when resources allow.

\section{Conclusion}\label{sec:conclusion}
We propose a cloud–device collaborative framework for CTR prediction under feature-asymmetric settings, where sensitive user attributes are restricted to the device. By separating cloud-side pre-ranking and device-side ranking, the framework enables personalization while preserving user privacy.
To address degraded candidate quality under limited cloud-accessible features, we introduce a diversity-aware pre-ranking mechanism that improves shortlist coverage, and a cloud-guided auxiliary learning strategy that enhances the lightweight on-device model without increasing inference cost. Together, these designs effectively bridge the gap between cloud-side efficiency and device-side personalization.
Extensive experiments on three public datasets demonstrate that the proposed framework achieves strong recommendation performance with low latency, while strictly keeping sensitive user features on-device.
For future work, we plan to explore how the pre-ranking diversity regularizer can serve as a mechanism to mitigate popularity bias, potentially surfacing "long-tail" items that are often overshadowed by dominant trends in standard ranking models. Finally, we will benchmark this architecture against fully cloud-based models to quantify the net benefits of decentralized, privacy-preserving CTR prediction.

\begin{acks}
This work was supported by the Insight Centre for Data Analytics (Grant No. SFI/12/RC/2289\_P2) and Huawei Ireland Research Center. 
We also acknowledge the computational facilities provided by the UCD Sonic High Performance Computing (HPC) cluster.
\end{acks}

\newpage
%\newpage
\bibliographystyle{ACM-Reference-Format}
\balance
\bibliography{reference}

\end{document}